\documentclass[11pt,twoside,a4paper]{article}
\usepackage{amsfonts}
\usepackage{amscd}
\usepackage{amssymb}
\usepackage{amsmath}
\usepackage{amstext}
\usepackage[a4paper]{geometry}
\usepackage[english]{babel}
\usepackage[T1]{fontenc}
\usepackage[latin1]{inputenc}
\usepackage{hyphenat}

\usepackage{graphicx}
\usepackage{color}
\usepackage[toc,page]{appendix}
\usepackage{hyperref}

\usepackage{floatrow}

\begin{document}
\title{Drifting states and synchronization induced chaos in 
       autonomous networks of excitable neurons}
\author{Rodrigo Echeveste, Claudius Gros}
\date{\today}
\maketitle

\begin{abstract}

The study of balanced networks of excitatory and inhibitory 
neurons has led to several open questions. On the one hand it 
is yet unclear whether the asynchronous state observed in the
brain is autonomously generated, or if it results from the
interplay between external drivings and internal dynamics.
It is also not known, which kind of network variabilities 
will lead to irregular spiking and which to synchronous 
firing states. Here we show how isolated networks of 
purely excitatory neurons generically show asynchronous 
firing whenever a minimal level of structural variability 
is present together with a refractory period.

Our autonomous networks are composed of excitable units, 
in the form of leaky integrators spiking only in response 
to driving currents, remaining otherwise quiet. For a 
non-uniform network, composed exclusively of excitatory 
neurons, we find a rich repertoire of self-induced dynamical 
states. We show in particular that asynchronous drifting states 
may be stabilized in purely excitatory networks whenever 
a refractory period is present. Other states found are 
either fully synchronized or mixed, containing both drifting 
and synchronized components. The individual neurons 
considered are excitable and hence do not dispose 
of intrinsic natural firing frequencies. An effective 
network-wide distribution of natural frequencies is 
however generated autonomously through self-consistent 
feedback loops. The asynchronous drifting state is, 
additionally, amenable to an analytic solution.

We find two types of asynchronous activity, with the individual
neurons spiking regularly in the pure drifting state,
albeit with a continuous distribution of firing frequencies.
The activity of the drifting component, however, becomes 
irregular in the mixed state, due to the periodic driving
of the synchronized component. We propose a new tool for
the study of chaos in spiking neural networks, which consists
of an analysis of the time series of pairs of consecutive 
interspike intervals. In this space, we show that a strange
attractor with a fractal dimension of about 1.8 is formed in 
the mentioned mixed state.
\end{abstract}

\section{Introduction}

The study of collective synchronization has attracted the 
attention of researchers across fields for now over half 
a century \cite{pikovsky2015dynamics, winfree1967biological, peskin1975mathematical, 
kuramoto1975self, buck1988synchronous}. Kuramoto's 
exactly solvable mean field model of coupled limit-cycles 
\cite{kuramoto1975self}, formulated originally by 
Winfree \cite{winfree1967biological}, has helped in this
context to establish the link between the distribution 
of natural frequencies and the degree of synchronization 
\cite{gros2010complex}. Moreover, the functional simplicity 
of this model, and other extensions, has permitted to 
analytically study the collective response of the system 
to external perturbations in the form of phase resets 
\cite{levnajic2010phase}.
Networks of phase coupled oscillators 
may show, in addition, attracting states corresponding to
limit cycles, heteroclinic networks, and chaotic phases
\cite{ashwin2007dynamics, dorfler2014synchronization}, with 
full, partial, or clustered synchrony 
\cite{golomb1992clustering}, or asynchronous behaviour 
\cite{abbott1993asynchronous}\\

Different degrees of collective synchronization may occur also in networks
of elements emitting signals not continuously, such as
limit-cycle oscillators, but via short-lived pulses
\cite{strogatz1993coupled,mirollo1990synchronization,abbott1993asynchronous}. 
Networks of pacemaker cells in the heart \cite{peskin1975mathematical}, 
for instance, synchronize with high precision, acting together 
as a robust macroscopic oscillator. Other well-known examples
are the simultaneous flashing of extended populations of 
southeast Asian fireflies \cite{buck1988synchronous, hanson1978comparative}
and the neuronal oscillations of cortical networks 
\cite{buzsaki2004neuronal}. In particular, the study 
of synchronization in the brain is of particular relevance
for the understanding of epileptic states, or seizures 
\cite{velazquez2007phase}.\\

The individual elements are usually modeled in this
context as integrate and fire units
\cite{kuramoto1991collective,izhikevich1999weakly}, where the 
evolution (in between pulses, flashes, or spikes) 
of a continuous internal state variable $V$ is governed 
by an equation of the type:
\begin{equation}
\tau \dot V \ =\ f(V) + I.
\label{eq_V_dot_generic}
\end{equation}
Here $\tau$ is the characteristic relaxation timescale of  
$V$, with $f$ representing the intrinsic dynamics of the unit, 
and $I$ the overall input (both from other units and from 
external stimuli). Whenever $V$ reaches a threshold value 
$V_{\theta}$, a pulse is emitted (the only information carried
to other units) and the internal variable is reset to
$V_{rest}$.\\

These units are usually classified either as oscillators or
as excitable units, depending on their intrinsic dynamics. 
The unit will fire periodically even in the absence of input
when $f(V) > 0$ ($\forall\, V\leq V_\theta$). Units of this kind 
are denoted \textit{pulse-coupled oscillators}. The unit is,
on the other hand, an \textit{excitable unit}, if an additional 
input is required to induce firing.\\

A natural frequency given by the inverse integration time of the 
autonomous dynamics exist in the case of pulse-coupled oscillators.
There is hence a preexisting, albeit discontinuous limit cycle, 
which is then perturbed by external inputs. One can hence use 
phase coupling methods to study networks of pulse 
coupled oscillators
\cite{mirollo1990synchronization,kuramoto1991collective,izhikevich1999weakly},
by establishing a map between the internal state variable 
$V$ and a periodic phase $\phi$ given by the state of the 
unit within its limit cycle. From this point of view
systems of pulse-coupled units share many properties with
sets of coupled Kuramoto-like oscillators \cite{kuramoto1975self}, 
albeit with generally more complex coupling functions 
\cite{izhikevich1999weakly}. For reviews and examples of 
synchronization in populations of coupled oscillators see 
\cite{strogatz2000kuramoto,dorfler2014synchronization}.\\

These assumptions break down for networks of coupled excitable 
units as the ones here described. 
In this case the units will remain silent without inputs 
from other elements of the system and there are no preexisting 
limit cycles and consequently also no preexisting natural 
frequencies (unlike \textit{rotators} \cite{sonnenschein2014cooperative}, 
which are defined in terms of a periodic phase variable, and 
a count with a natural frequency). The firing rate depends hence exclusively on the 
amount of input received. The overall system activity will
therefore forcefully either die out or be sustained collectively 
in a self-organized fashion \cite{gros2010complex}. The respectively 
generated spiking frequencies for a given mean network activity 
could be considered in this context 
as self-generated natural frequencies.\\

The study of pulse coupled excitable units is of particular 
relevance within the neurosciences, where neurons are 
often modeled as spike emitting units that continuously integrate 
the input they receive from other cells \cite{burkitt2006review}.
The proposal \cite{shadlen1994noise, amit1997model}, and later 
the empirical observation that excitatory and inhibitory inputs
to cortical neurons are closely matched in time 
\cite{sanchez2000cellular, haider2006neocortical}, has led 
researchers to focus on dynamical states (asynchronous states
in particular) in networks characterized by a balance between
excitation and inhibition
\cite{van1996chaos, kumar2008high, vogels2005signal, stefanescu2008low,hansel2001existence,abbott1993asynchronous}.
This balance (E/I balance) is generally presumped to be
an essential condition for the stability of states showing
irregular spiking, such as the one arising in balanced 
networks of integrate and fire neurons \cite{brunel2000dynamics}.
The type of connectivity usually employed in network studies however, 
is either global, or local consisting of either repeated patterns, 
or random connections drawn from identical distributions 
\cite{kuramoto2002coexistence,alonso2011average,abrams2004chimera, ashwin2007dynamics}. 
Our results show, however, that only a minimal level of structural
variability is necessary for excitatory networks to display wide 
varieties of dynamical states, including stable autonomous irregular 
spiking. We believe that these studies are not only interesting on 
their own because of the richness of dynamical states, but also 
provide valuable insight into the role of inhibition.\\

Alternatively, one could have built networks of excitatory 
neurons with high variability in the connection parameters, 
reproducing realistic connectivity distributions, such as 
those found in the brain. 
The large number of parameters involved would make 
it however difficult to fully characterize the system from 
a dynamical systems point of view, the approach taken here. 
An exhaustive phase-space study would also become intractable. 
We did hence restrict ourselves in the present work to a 
scenario of minimal variability, as given by a network of 
globally coupled excitatory neurons, where the coupling 
strength of each neuron to the mean field is non-uniform. 
Our key result is that stable irregular 
spiking states emerge even when only a minimal level of 
variability is present at a network level.\\

Another point we would like to stress here is that
asynchronous firing states may be stabilized in the
absence of external inputs. In the case here studied, there is an
additional `difficutlty' to the problem, in the sense that the 
pulse-coupled units considered are in excitable 
states, remaining quiet without sufficient drive from the 
other units in the network. The observed sustained 
asynchronous activity is hence self-organized.\\
 
We characterize how the features of the network dynamics depend 
on the coupling properties of the network and, in particular, 
we explore the possibility of chaos in the
 here studied case of excitable units, when partial synchrony 
is present, since this link has already been established in the case 
of coupled oscillators with a distribution 
of natural frequencies \cite{miritello2009central}, while other studies 
had also shown how stable chaos emerges in inhibitory networks of 
homogeneous connection statistics \cite{angulo2014stable} \\

\section{The Model}

In the current work we study the properties of the 
self-induced stationary dynamical states in autonomous 
networks of excitable integrate-and-fire neurons.
The neurons considered 
are characterized by a continuous state variable $V$ 
(as in Eq.~(\ref{eq_V_dot_generic})), representing the 
membrane potential, and a discrete state variable $y$ 
that indicates whether the neuron fires a spike ($y=1$) 
or not ($y=0$) at a particular point in time. More precisely, 
we will work here with a conductance based (COBA) integrate-and-fire 
(IF) model as employed in \cite{vogels2005signal} (here
however without inhibitory neurons), in which the evolution 
of each neuron $i$ in the system is described by:
\begin{equation}
\tau\dot V_i \ =\ \left(V_{rest}-V_i\right) + g_i \left(E_{ex}-V_i\right),
\label{eq_V_dot}
\end{equation}
where $E_{ex} = 0\,\mathrm{mV}$ represents 
the excitatory reversal potential and $\tau = 20\,\mathrm{ms}$ 
is the membrane time constant. Whenever the membrane potential 
reaches the threshold $V_{\theta} = -50\,\mathrm{mV}$, the 
discrete state of the neuron is set to $y_i= 1$ for the duration of 
the spike. The voltage is reset, in
addition, to its resting value of $V_{rest} = -60\,\mathrm{mV}$,
where it remains fixed for a refractory period of 
$t_{ref} = 5\,\mathrm{ms}$. Eq.~(\ref{eq_V_dot}) is not computed 
during the refractory period. Except for the times of spike 
occurrences, the discrete state of the neuron remains $y_i=0$ 
(no spike).\\

The conductance $g_i$ in (\ref{eq_V_dot}) integrates the 
influence of the time series of presynaptic spikes, decaying 
on the other side in absence of inputs:
\begin{equation}
\tau_{ex}\; \dot g_i \ =\ -g_i,
\label{eq_g_dot}
\end{equation}
where $\tau_{ex} = 5\,\mathrm{ms}$ is the conductance 
time constant. Incoming spikes from the $N-1$ other 
neurons  produce an increase of the conductance 
$g_i \longrightarrow g_i + \Delta g_i$, with:
\begin{equation}
\Delta g_i \ =\ \frac{K_i}{N-1}\;\sum_{j\neq i}  w_{ij} \, y_j.
\label{eq_g}
\end{equation}
Here the synaptic weights $w_{ij}$ represent the intensity 
of the connection between the presynaptic neuron $j$ and the 
postsynaptic neuron $i$. We will generally employ normalized
synaptic matrices with $\sum_j w_{ij}/(N-1)=1$. In this way
we can scale the overall strength of the incoming connections 
via $K_i$, retaining at the same time the structure of the 
connectivity matrix.

\subsection{Global couplings}

Different connectivity structures are usually employed in 
the study of coupled oscillators, ranging from purely 
local rules to global couplings
\cite{kuramoto2002coexistence,alonso2011average,abrams2004chimera, ashwin2007dynamics}.
We start here by employing a global coupling structure, 
where each neuron is coupled to the overall firing activity 
of the system:
\begin{equation}
w_{ij} = 1 \quad \forall \quad i\neq j, \qquad w_{ii} = 0,
\label{eq_w}
\end{equation}
which corresponds to a uniform connectivity matrix without 
self coupling. All couplings are excitatory. The update 
rule (\ref{eq_g}) for the conductance upon presynaptic 
spiking then take the form:
\begin{equation}
\Delta g_i \ =\ K_i \frac{\sum_{j\neq i} y_j}{N-1} \ =\ K_i \, \bar{r},
\qquad\quad
\bar{r} = \frac{\sum_{j\neq i} y_j}{N-1},
\label{eq_g2}
\end{equation}
where $\bar{r}=\bar{r}(t)$ represents the time-dependent mean 
field of the network, viz the average over all firing activities.
$\bar{r}$ is hence equivalent to the mean field present
in the Kuramoto model \cite{kuramoto1975self}, resulting
in a global coupling function as an aggregation of local 
couplings. With our choice (\ref{eq_w}) for the coupling 
matrix the individual excitable units may be viewed,
whenever the mean field $\bar{r}$ is strong enough, as
oscillators emitting periodic spikes with an `effective'
natural frequency determined by the afferent coupling 
strength $K_i$. The resulting neural activities determine
in turn the mean field $\bar{r}(t)$.

\subsection{Coupling strength distribution}

We are interested in studying networks with non-uniform $K_i$, 
We mostly consider here the case of equidistant $K_i$, defined by:
\begin{equation}
K_i \ = \ \bar{K} - \Delta K + \frac{2\Delta K}{N-1} \left(i-1\right),
\qquad\quad i=1,\dots,N
\label{eq_Ki}
\end{equation}
for the $N$ neurons, where $\bar{K}$ represents the mean 
scaling parameter,  and $\Delta K$, the maximal distance to the 
mean. It is possible, alternatively, 
to use a flat distribution with the $K_i$ drawn from an interval
$[\bar{K}-\Delta K,\bar{K}+\Delta K]$ around
the mean $\bar{K}$. For large systems there is no
discernible difference, as we have tested, between using
equidistant afferent coupling strengths $K_i$ and drawing
them randomly from a flat distribution.

\section{Results}

Several aspects of our model, in particular the asynchronous 
drifting state, can be investigated analytically as a 
consequence of the global coupling structure (\ref{eq_w}),
as shown in Sect.~\ref{subsec_drifting}. All further
results are obtained from numerical simulations, for which,
if not otherwise stated, a timestep of $0.01\,\mbox{ms}$ has 
been used. We have also set the spike duration to one time-step, 
although these two parameters can be modified separately if 
desired, with our results not depending on the choice 
of the time-step, while the spike width does introduce minor
quantitative changes to the results, as later discussed.\\

\subsection{Stationary mean-field solution for the drifting state}
\label{subsec_drifting}

As a first approach we compute the response of a neuron 
with coupling constant $K_i$ to a stationary mean field 
$\bar{r}$, as defined by Eq.~(\ref{eq_g2}), representing 
the average firing rate of spikes (per second) of the 
network. This is actually the situation present in the
asynchronous drifting state, for which the firing
rates of the individual units are incommensurate.
With $\bar{r}$ being constant we can combine 
the update rules (\ref{eq_g_dot}) and (\ref{eq_g})
for the conductances $g_i$ to
\begin{equation}
\tau_{ex}\; \dot g_i \ =\ -g_i + \tau_{ex} K_i \bar{r},
\qquad\quad
 g_i^*  \ =\ \tau_{ex} K_i \bar{r},
\label{eq_g_dot_MF}
\end{equation}
where we have denoted with $g_i^*$ the steady-state conductance.
With the individual conductance becoming a constant
we may also integrate the evolution equation (\ref{eq_V_dot})\
for the membrane potential,
\begin{equation}
\tau\dot V_i \ =\ \left(V_{rest}-V_i\right) + 
                  \tau_{ex} K_i \bar{r} \left(E_{ex}-V_i\right),
\label{eq_V_dot_MF}
\end{equation}
obtaining the time $t^*_i$ it takes for the membrane potential
$V_i$ to  reach the threshold $V_\theta$, when starting from
the resting potential $V_{rest}$:
\begin{equation}
t^*_i \ =\ - \frac{1}{B_i} 
log\left( \frac{B_i V_{\theta} -A_i}{B_i V_{rest} -A_i}\right),
\label{eq_t_star_MF}
\end{equation}
with:
\begin{equation}
A_i \ =\ \frac{V_{rest} + \tau_{ex} K_i \bar{r} E_{ex}}{\tau}, 
\qquad  B_i \ =\ \frac{1 + \tau_{ex} K_i \bar{r}}{\tau}.
\label{A_B_MF_E_only}
\end{equation}
We note, that the threshold potential $V_\theta$ is only
reached, if $dV_i/dt>0$ for all $V_i\le V_\theta$. For
the $t_i^*$ to be finite we hence have (from Eq.~(\ref{eq_V_dot_MF}))
\begin{equation}
K_i \bar{r} > \frac{1}{\tau_{ex}} 
\left( \frac{V_\theta - V_{rest}}{ E_{ex} - V_\theta} \right).
\label{condition}
\end{equation}
The spiking frequency is $r_i =T_i^ {-1}$, with the intervals
$T_i$ between consecutive spikes given by $T_i = t^*_i + t_{ref}$, 
when (\ref{condition}) is satisfied. Otherwise the neuron does
not fire. The mean field $\bar{r}$ is defined as
the average firing frequency 
\begin{equation}
\bar{r} \ =\ \left< r_i \right>  \ =\ \left< \frac{1}{t^*_i + t_{ref}} \right>
\label{eq_r_self_cons_MF}
\end{equation}
of the neurons. Eqs.~(\ref{eq_r_self_cons_MF}) and 
(\ref{eq_t_star_MF}) describe the asynchronous drifting 
state in the thermodynamic limit $N\to\infty$. We denote 
this self-consistency condition for $\bar{r}$ the stationary 
mean-field (SMF) solution.

\begin{figure}[!t]
\begin{center}
\includegraphics[width=\textwidth]{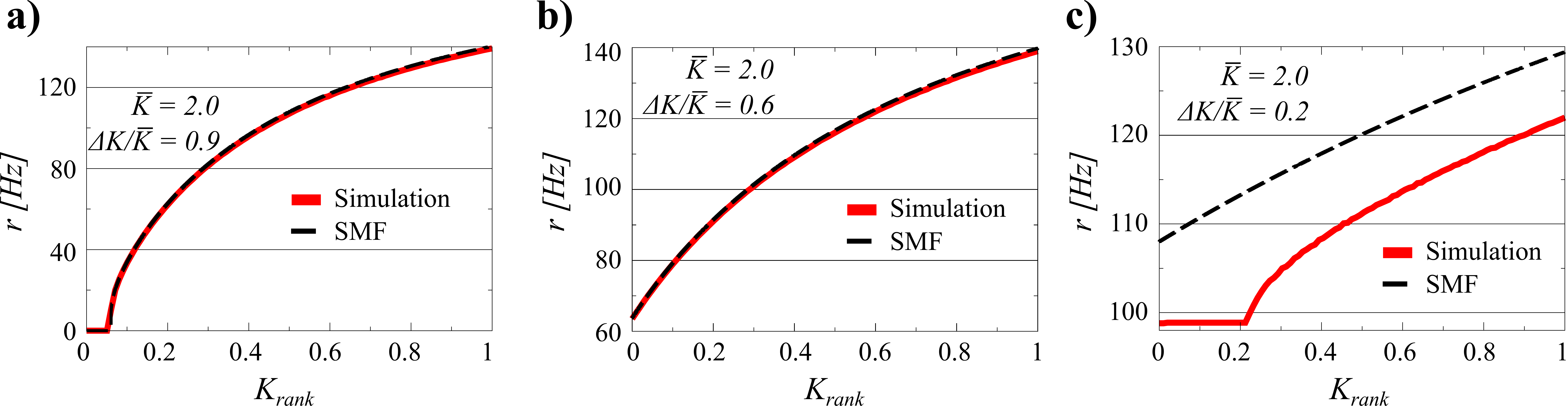}
\end{center}
\caption{The firing rates of all $i=1,\dots,N$ neurons, 
as a function of the relative rank $K_{rank}=i/N$ 
of the individual neurons ($N=100$). The coupling 
matrix is uniform (see Eq.~(\ref{eq_w})) and the 
afferent coupling strength $K_i$ uniformly distributed 
between $\bar{K}\pm \Delta K$; with $\bar{K} = 2.0$ 
and $\Delta K/\bar{K} = 0.9/0.6/0.2$ 
(left/middle/right panel). The full red lines
denote the results obtained by solving numerically
Eqs.~(\ref{eq_V_dot}) and (\ref{eq_g_dot}), and
the dashed lines the stationary mean field 
solution (SMF, Eq.~\ref{eq_r_self_cons_MF}). 
}
\label{fig_examples}
\end{figure}

\subsection{Numerical simulations}

We studied our model, as defined by Eqs.~(\ref{eq_V_dot}) 
and (\ref{eq_g_dot}), numerically for networks with 
typically $N=100$ neurons, a uniform coupling matrix 
(see Eq.~(\ref{eq_w})) and coupling parameters $\bar{K}$ 
and $\Delta K$ given by Eq.~(\ref{eq_Ki}). We did not
find qualitative changes when scaling the size of the network 
up to $N = 400$ for testing purposes (and neither with
down-scaling), see Fig.~\ref{fig_robustness}. 
Random initial conditions where used. 
The network-wide distribution of firing rates is 
computed after the system settles to a 
dynamical equilibrium.\\

Three examples, for $\bar{K} = 2.0$ and 
$\Delta K/\bar{K} = 0.9$, $0.6$, and $0.2$, 
of firing-rate distributions are presented in 
Fig.~\ref{fig_examples} in comparison with the analytic
results obtained from the stationary mean field 
approach (\textit{SMF}), as given by Eq.~(\ref{eq_r_self_cons_MF}).
The presence or absence of synchrony is directly visible.
In all of the three parameter settings presented 
in Fig.~\ref{fig_examples} there is a drifting component, 
characterized by a set of neurons with a continuum of 
frequencies. These neurons fire asynchronously,
generating a constant contribution to the collective 
mean field.\\

The plateau present in the case $\Delta K/\bar{K} = 0.2$, 
corresponds, on the other hand, to a set of neurons 
firing with identical frequencies and hence synchronously.
Neurons firing synchronously will do so however with
finite pairwise phase lags, with the reason being
the modulation of the common mean field $\bar{r}$ 
through the distinct afferent coupling strengths $K_i$.
We note that the stationary mean-field theory 
(\ref{eq_r_self_cons_MF}) holds, as expected, for 
drifting states, but not when synchronized clusters 
of neurons are present.\\

\begin{figure}[t!]
\begin{center}
\includegraphics[width=\textwidth]{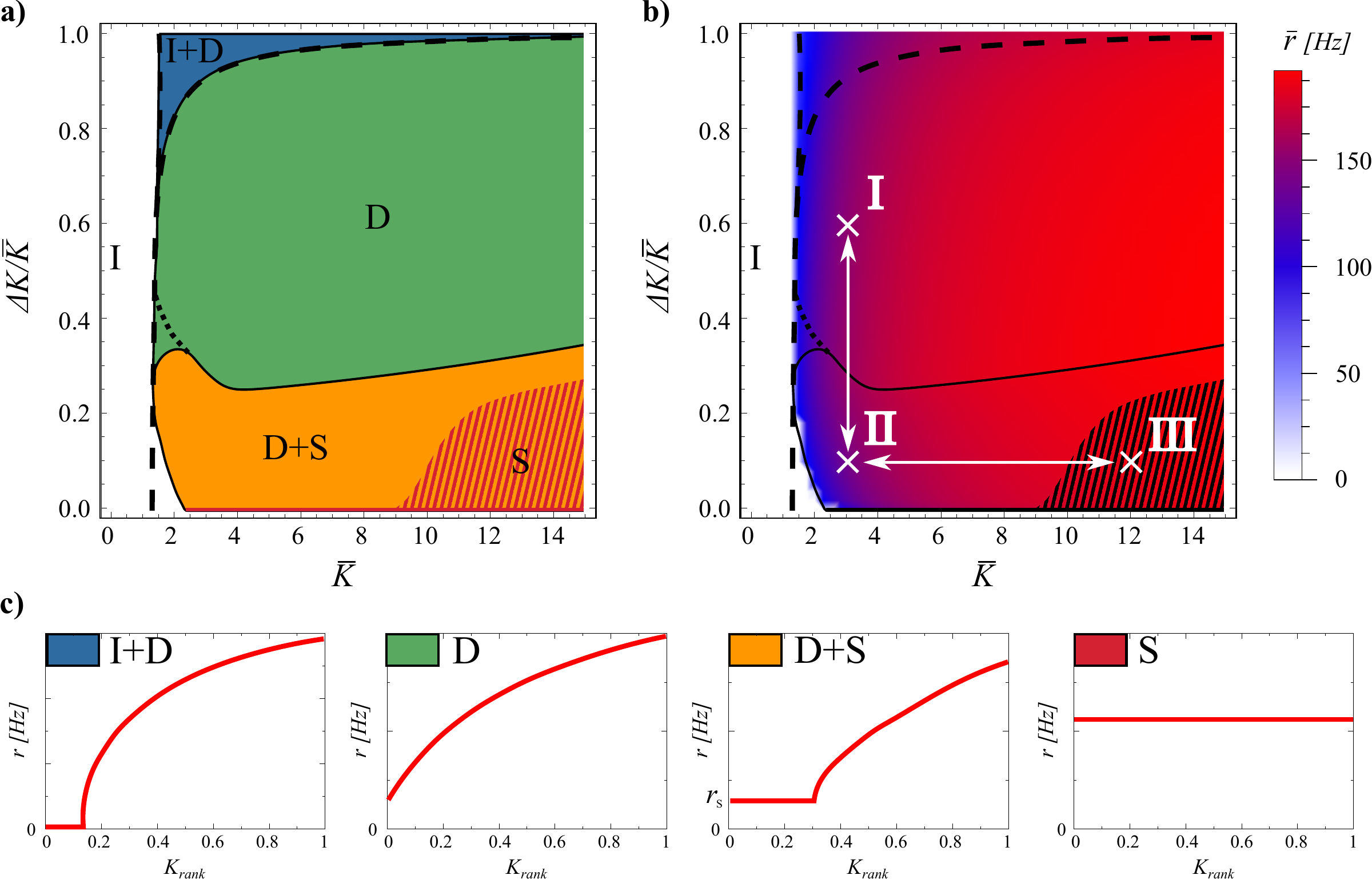}
\end{center}
\caption{The phase diagram, as obtained for a network of 
$N=100$ neurons evolving according to Eqs.~(\ref{eq_V_dot}) 
and (\ref{eq_g_dot}). The network matrix is flat, see 
Eq.~(\ref{eq_w}). Full and partially inactive (I), drifting 
(D) and synchronized states (S) are found as a function of 
the coupling parameters $\bar{K}$ and $\Delta K$ (Eq.~\ref{eq_Ki}).
\textbf{a)} The dashed lines represent the phase transition 
lines as predicted by the stationary mean field approximation. 
(\ref{eq_r_self_cons_MF}). The shaded region indicates 
the coexistence of attracting states S and S+D. \textbf{b)} The average 
firing rate of the network. In black the phase boundaries and
in white the two adiabatic paths used in Fig.~\ref{fig_adiabatics}.
\textbf{c)} Examples of the four active dynamical states found. 
As in Fig.~\ref{fig_examples}.
}
\label{fig_phase_space}
\end{figure}

In Fig.~\ref{fig_phase_space} we systematically explore 
the phase space as a function of $\bar{K}$ and 
$\Delta K$. For labeling the distinct phases we 
use the notation

\begin{tabular}{rcl}
I  &:&inactive,\\
I+D&:&partially inactive and drifting,\\
D  &:&fully drifting (asynchronous),\\
D+S&:&mixed, containing both drifting and synchronized components, and \\
S  &:&fully synchronized.
\end{tabular}\\
Examples of the rate distributions present in the
individual phases are presented in 
Fig.~\ref{fig_phase_space}~\textbf{c)}.\\

The phase diagram is presented in
Fig.~\ref{fig_phase_space}~\textbf{a)}.
The activity dies out for a low mean 
connectivity strength $\bar{K}$, but not for 
larger $\bar{K}$. Partial synchronization is present 
when both $\bar{K}$ and the variance $\Delta K$ are small, 
taking over completely for larger values of $\bar{K}$
and small $\Delta K$. The phase space is otherwise 
dominated by a fully drifting state. The network
average $\bar{r}$ of the neural firing rates, given in
Fig.~\ref{fig_phase_space}~\textbf{b)}, drops only
close to the transition to the inactive state I,
showing otherwise no discernible features at the
phase boundaries.\\

The dashed lines in Fig.~\ref{fig_phase_space}~\textbf{a)} 
and \textbf{b)} represent the transitions between the 
inactive state I and active state I+D, and between states I+D 
and D, as predicted by the stationary mean field 
approximation (\ref{eq_r_self_cons_MF}), which becomes
exact in the thermodynamic limit. The shaded region 
in these plots indicates the co-existence of 
attracting states S and S+D. As a note, we found 
that the location of this shaded region depends on 
the spike width, shifting to higher $\bar{K}$ values 
for narrower spikes. While real spikes in neurons 
have a finite width, we note from a dynamical systems 
point of view, that this region would most likely 
vanish in the limit of delta spikes.\\

For a stable (non-trivial) attractor to arise in a 
network composed only of excitatory neurons, 
some limitation mechanism needs to be at play. Otherwise 
one observes a bifurcation phenomenon, similar to that 
of branching problems, in which only a critical network 
in the thermodynamic limit could be stable \cite{gros2010complex}. 
In this case, 
the limiting factor is the refractory period. 
Refractoriness prevents neurons from firing continuously, 
and prevents the system activity from exploding. Interestingly, 
this does not mean that the neurons will fire at the 
maximal rate of $1/t_{ref}$ which would correspond in 
this case to $200\,\mathrm{Hz}$. The existence of this 
refractory period allows for self-organized states 
with frequencies even well bellow this limit, as seen 
in Fig.~\ref{fig_phase_space}~\textbf{b)}. We have tested 
these claims numerically by setting $t_{ref} = 0$, observing that 
the neural activity either dies out or the neurons fire 
continuously.\\

In order to study the phase transitions 
between states D and D+S and between D+S and S, we will resort 
in the following section to adiabatic paths in phase space 
crossing these lines.\\

\begin{figure}[!t]
\begin{center}
\includegraphics[width=0.9\textwidth]{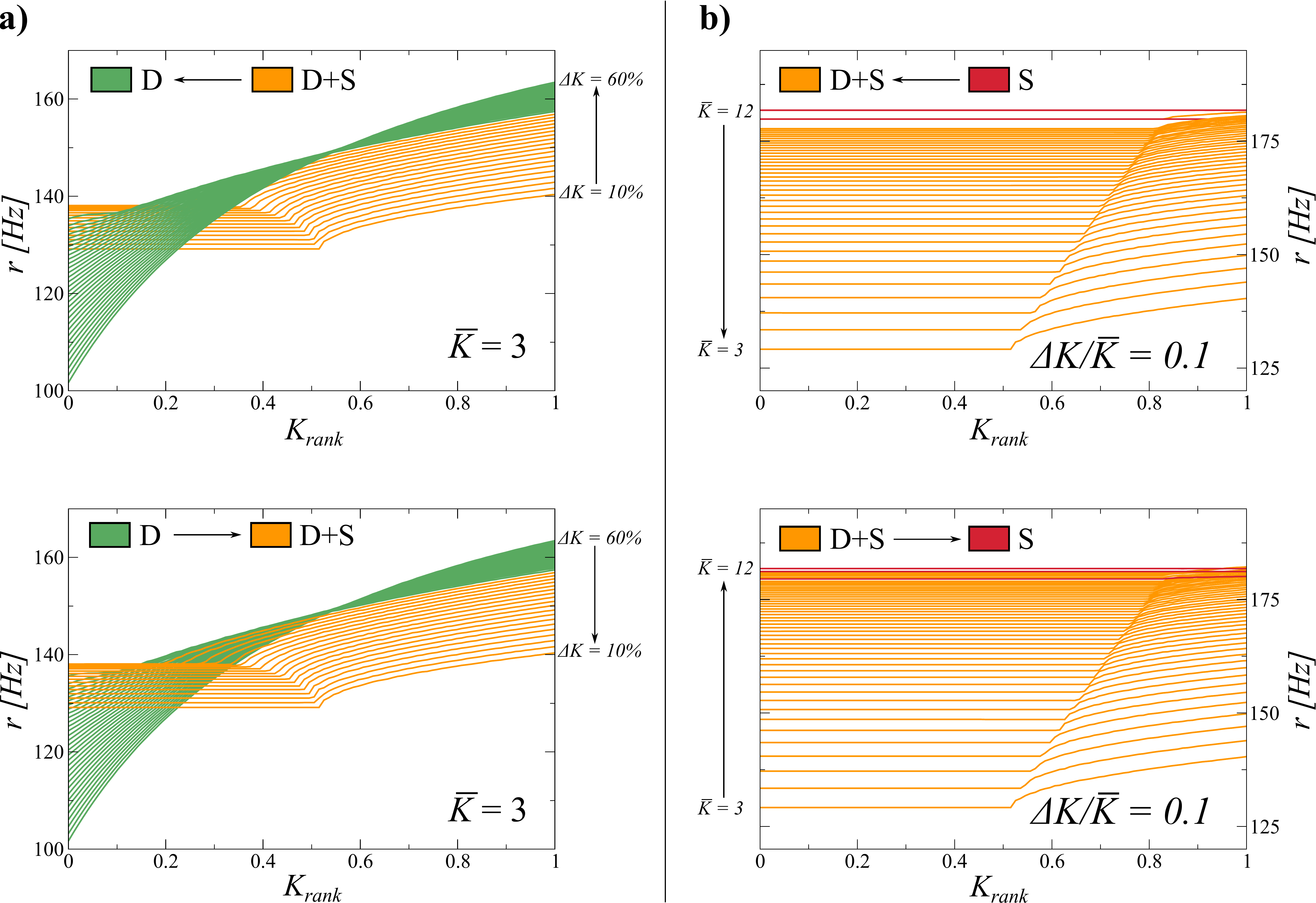}
\end{center}
\caption{Study of the transitions between fully drifting 
$D$ and partially drifting and synchronized $D+S$
phase (left panels), and between $D+S$ and the fully
synchronized $S$ state (right panels). For the full
phase diagram see Fig.~\ref{fig_phase_space}. The coupling 
parameters $\bar{K}$ and $\Delta K$ (Eq.~\ref{eq_Ki}), 
are modified on times scales much slower than the intrinsic
dynamics . For the two adiabatic paths considered, 
each crossing a phase transition line, the evolution of 
the firing rate distribution is computed in several windows 
and shown.
\textbf{a)} $\bar{K} = 3.0$ is kept constant and 
$\Delta K/\bar{K}$ varied between $0.1$ and $0.6$ (from
$D$ to $D+S$, indicated as I$\leftrightarrow$II in Fig.~\ref{fig_adiabatics}).
\textbf{b)} $\Delta K/\bar{K}=0.1$ is kept constant,
varying  $\bar{K}$ between $3$ and $12$ (from $D+S$ to $S$,
indicated as II$\leftrightarrow$III in Fig.~\ref{fig_adiabatics}).
}
\label{fig_adiabatics}
\end{figure}

\subsubsection{Adiabatic parameter evolution}

Here we study the nature of the phase transitions between 
different dynamical states in Fig.~\ref{fig_phase_space}. To do so, 
we resort to adiabatic trajectories in phase space, crossing these 
lines. Beginning in a given phase we modify the 
coupling parameters $\bar{K}$ and $\Delta K$ (Eq.~\ref{eq_Ki}), 
on a timescale much slower than that of the network dynamics. 
Along these trajectories, we then freeze the system in a 
number of windows in which we compute the rate distribution 
as a function of the $K_{rank}$ (see Fig.~\ref{fig_examples}).
During these observation windows the parameters do not change.
In this way, we can follow how the rate distribution varies 
across the observed phase transitions. The results are presented 
in Fig.~\ref{fig_adiabatics}.\\

We observe that the emergence of synchronized clusters,
the transition D$\to$(D+S), is completely reversible. We
believe this transition to be of second order and that
the small discontinuity in the respective firing rate
distributions observed in Fig.~\ref{fig_adiabatics}{\bf a)} 
are due to finite-size effects. The time to reach the
stationary state diverges, additionally, close to the
transition, making it difficult to resolve the locus
with high accuracy.\\

The disappearance of a subset of drifting neurons, the 
transition S$\to$(D+S) is, on the other hand, not 
reversible. In this case, when $\bar{K}$ is reduced, 
the system tends to get stuck in metastable attractors 
in the $S$ phase, producing irregular jumps in the rate 
distributions. Furthermore, when we increase $\bar{K}$, 
we observe that the system jumps back and forth between 
states $D+S$ and $S$ in the vicinity of the phase transition, 
indicating that both states may coexist as metastable 
attractors close to the transition. We note that a
similar metastability has been observed in partially
synchronized phase of the Kuramoto model 
\cite{miritello2009central}.\\

\subsubsection{Time structure}\label{sec_time_structure}

\begin{figure}[!t]
\begin{center}
\includegraphics[width=\textwidth]{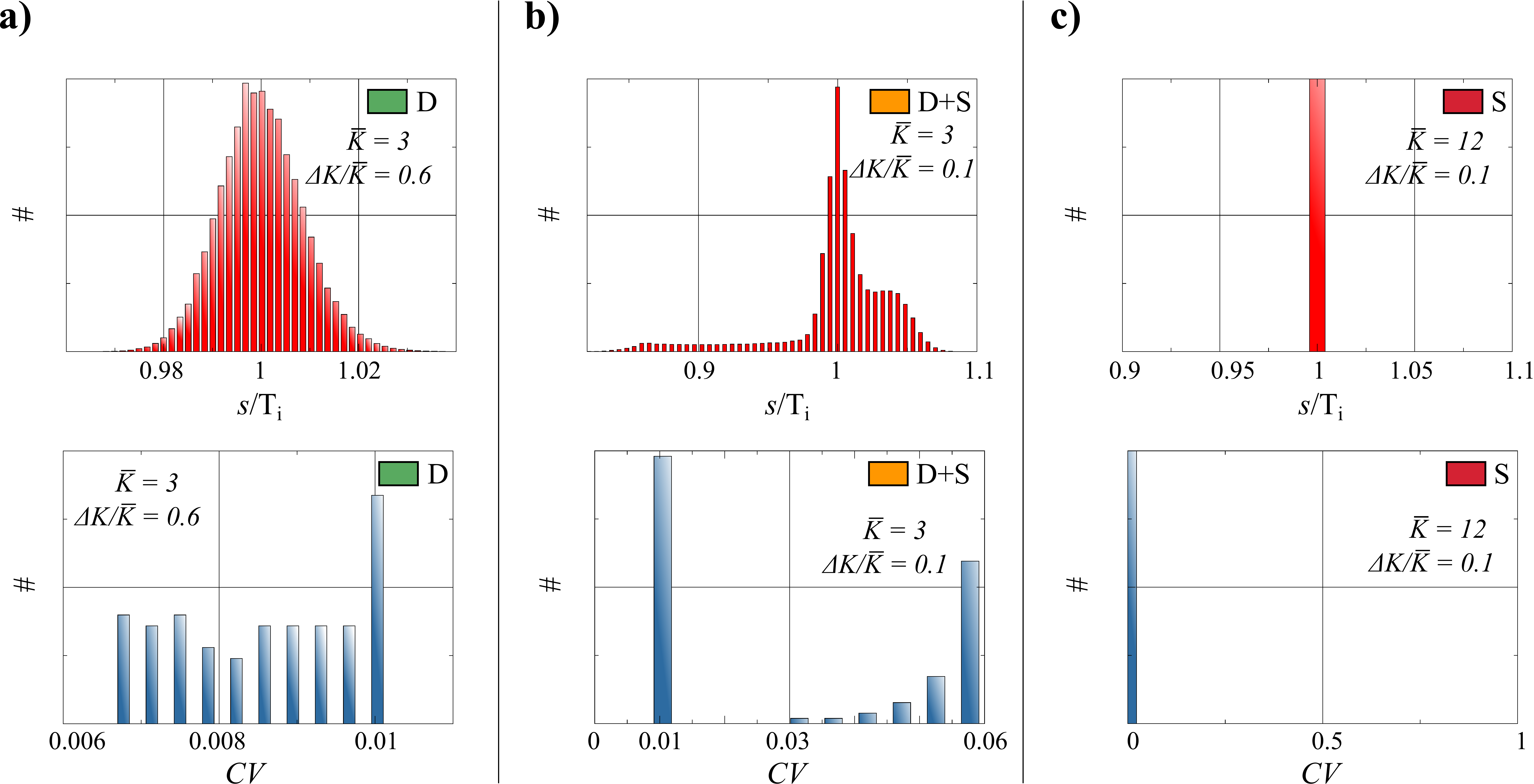}
\end{center}
\caption{Top: Histograms of interspike interval (ISI), denoted as $s$, 
normalized by the average period, for three parameter 
configurations. Bottom: Histograms of the coefficient of variation $CV$, 
as defined by Eq.~\ref{eq_fano}. Parameters (Eq.~\ref{eq_Ki}) 
and state (as defined in Fig.~\ref{fig_phase_space}), 
for both Top and Bottom:
\textbf{a)} $\bar{K} = 3.0$, $\Delta K/\bar{K} = 0.6$, state: D.
\textbf{b)} $\bar{K} = 3.0$, $\Delta K/\bar{K} = 0.1$, state: D+S.
\textbf{c)} $\bar{K} = 12.0$, $\Delta K/\bar{K} = 0.1$, state: S.
}
\label{fig_fano}
\end{figure}

In networks of spiking neurons, it is essential to 
characterize not only the rate distribution of the 
system, but also the neurons' interspike-time statistics 
\cite{lindner2004interspike,perkel1967neuronal_I,perkel1967neuronal_II,
farkhooi2009serial, chacron2004noise}. In this case, 
we have computed the distribution
$p_i(s)$ of the interspike intervals $s$ (\textit{ISI}) 
of the individual neurons respectively for full and 
partial drifting and synchronized states. The
distribution of inter-spike intervals 
in Fig.~\ref{fig_fano} shows the network average 
of the $p_i(s)$, normalized individually with
respect to the average $T_i =\int s\, p_i(s)ds$
spiking intervals. 
\begin{itemize}
\item D~: The input received by
   a given neuron $i$ tends to a constant,
   as discussed in Sect.~\ref{subsec_drifting},
   in the thermodynamic limit $N\to\infty$. The 
   small but finite width of the ISI for the fully 
   drifting state D evident in Fig.~\ref{fig_fano} 
   is hence a finite-size effect.
\item D+S~: The input received for drifting neuron $i$
   in a state where other neurons form a synchronized
   subcluster is intrinsically periodic in time and
   the resulting $p_i(s)$ non-trivial, as evident
   in Fig.~\ref{fig_fano}.
\item S~: $p_i(s)$ is a delta function for all
   neurons in the fully synchronized state, with 
   identical inter-spike intervals $T_i$.
\end{itemize}

As a frequently used measure of the regularity 
of a time distribution we have included
in Fig.~\ref{fig_fano}
the coefficient of variation ($CV$),
\begin{equation}
CV_i = \frac{\sigma_i}{T_i},  \qquad\quad
T_i =\int s\, p_i(s)ds, \qquad\quad
\sigma_i^2 =\int (s-T_i)^2p_i(s)ds~.
\label{eq_fano}
\end{equation}

Of interest here are the finite $CV$s of the
drifting units in the D+S state, which are considerably 
larger than the $CV$s of the drifting neurons when no
synchronized component is present. This 
phenomenon is a consequence of the interplay between 
the periodic driving of the drifting neurons by the 
synchronized subcluster in the D+S state, where the 
driving frequency will in general be in mismatch with 
the effective, self-organized natural frequency of the
drifting neurons. The firing of a drifting neuron
is hence irregular in the mixed D+S state, becoming
however regular in the absence of synchronized
drivings.  

\begin{figure}[!t]
\begin{center}
\includegraphics[width= 0.7 \textwidth]{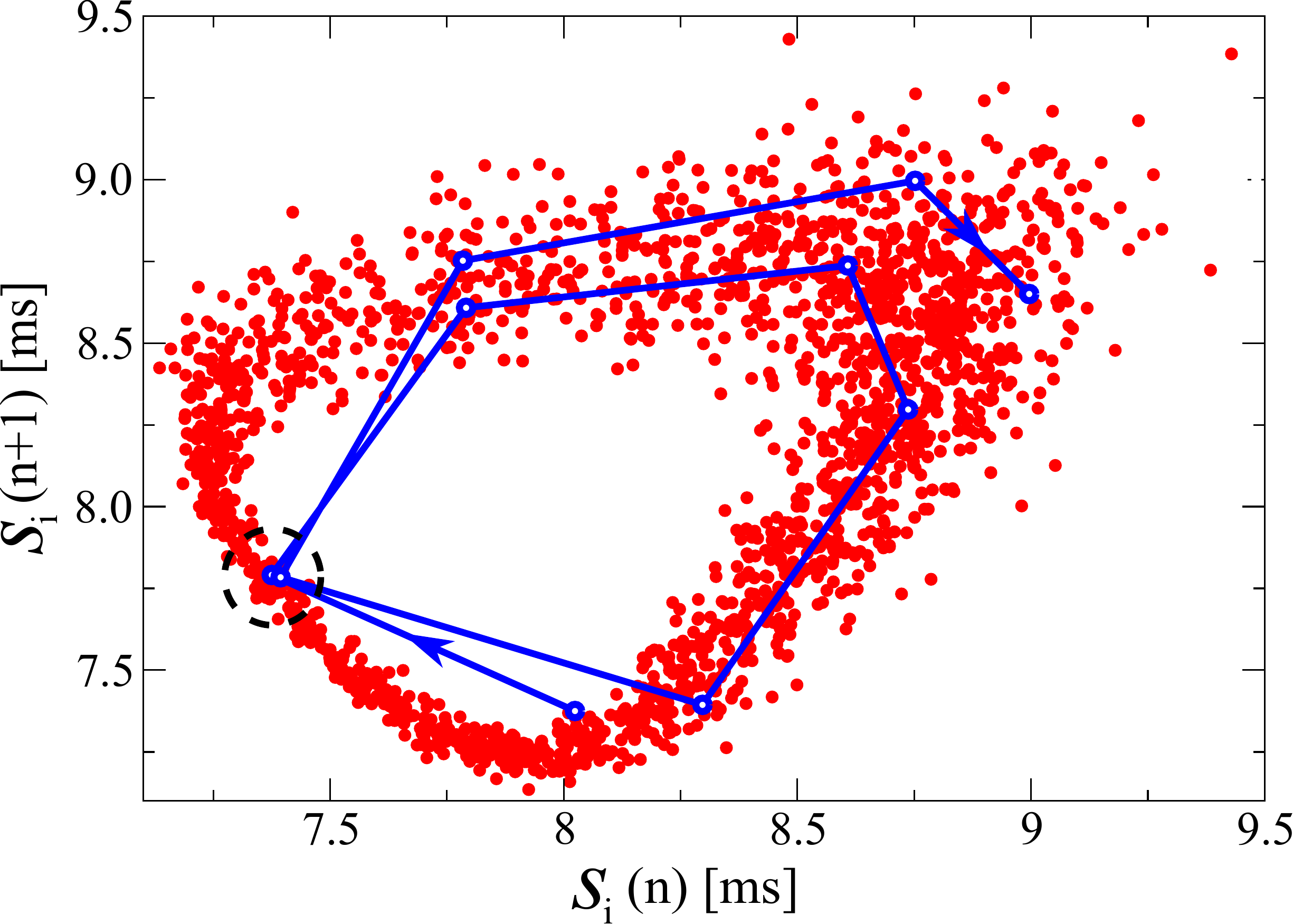}
\end{center}
\caption{Pairs of consecutive interspike intervals $s$ plotted 
against each other, for one of the drifting neurons in the D+S state, 
corresponding to the third example of Fig.~\ref{fig_examples} ($\bar{K} = 2.0$ 
and $\Delta K/\bar{K} = 0.2$). The plots are qualitatively similar for 
all drifting neurons in this state. The qualitative features of the plot 
are the same for any of the drifting neurons in this state. In red, 
each point represents a pair $\left[s_i(n),s_i(n+1)\right]$ where 
$n$ denotes the spike number. In blue, we follow a 
representative segment of the trajectory. The system does not 
appear to follow a limit cycle, and these preliminary results 
would suggest the presence of chaos in the D+S state, consistent 
with studies of chaos in periodically driven oscillators 
\cite{d1982chaotic}. Indeed, 
if one looks at the two close points within the dashed circle, we 
observe how an initially small distance between them, rapidly grows 
in a few iterations steps, indicating a positive eigenvalue.
For this simulation we have used a time-step $dt = 0.001\,\mathrm{ms}$, 
to improve the resolution of the points in the plot. We have 
evaluated the time the neuron needs to circle the attractor, 
finding it to be of the order of $\sim 5.3\,$ spikes. Other
drifting neurons take slightly longer or shorter. In 
Fig.~\ref{fig_fractal_dim}, we compute the fractal dimension 
of the here shown attractor.
}
\label{fig_chaos}
\end{figure}
\subsubsection{Self induced chaos}\label{sec_chaos}

The high variability of the spiking intervals observed
in the mixed state, as presented in Fig.~\ref{fig_fano},
indicates that the firing may have a chaotic component
in the mixed state and hence positive Lyapunov exponents.
\cite{gros2010complex}.\\

Alternatively to a numerical evaluation of the Lyapunov exponents 
(a demanding task for large networks of spiking neurons), 
a somewhat more direct understanding of the chaotic state 
can be obtained by studying the relation between consecutive 
spike intervals. In Fig.~\ref{fig_chaos} we plot for this
purpose a time series of 2000 consecutive interspike intervals 
$\left[s_i(n),s_i(n+1)\right]$ (corresponding to about
$17\,\mathrm{sec}$ in real time), for one of the drifting 
neurons in the D+S state (with the parameters of the 
third example of Fig.~\ref{fig_examples}: $\bar{K} = 2.0$ 
and $\Delta K/\bar{K} = 0.2$). We note that the spiking
would be regular, viz $s_i(n)$ constant, for all neurons either
in the fully drifting state (D) or in the fully synchronized
state (S). The plot of consecutive spike intervals
presented in Fig.~\ref{fig_chaos} can be viewed as
a poor man's approximation to Takens' embedding
theorem \cite{takens1981detecting}, which states
that a chaotic attractor arising in a $d$-dimensional
phase space (in our case $d=2N$) can be reconstructed 
by the series of $d$-tuples of time events of a single 
variable.

With a blue line we follow in Fig.~\ref{fig_chaos} a
representative segment of the trajectory, which jumps
irregularly. A first indication that the attractor in
Fig.~\ref{fig_chaos} may be chaotic comes from the
observation that the trajectory does not seem to 
settle (within the observation window) within a limit 
cycle. The time series of consecutive spike-interval
pairs will nevertheless approach any given previous 
pair $\left[s_i(n),s_i(n+1)\right]$ arbitrarily close,
a consequence of the generic ergodicity of attracting 
sets \cite{gros2010complex}. One of these close 
re-encounters occurs in Fig.~\ref{fig_chaos} 
near the center of the dashed circle, with the
trajectory diverging again after the close 
re-encounter. This divergence indicates the presence
of positive Lyapunov exponents.\\

\begin{figure}[!t]
\begin{center}
\includegraphics[width=0.6\textwidth]{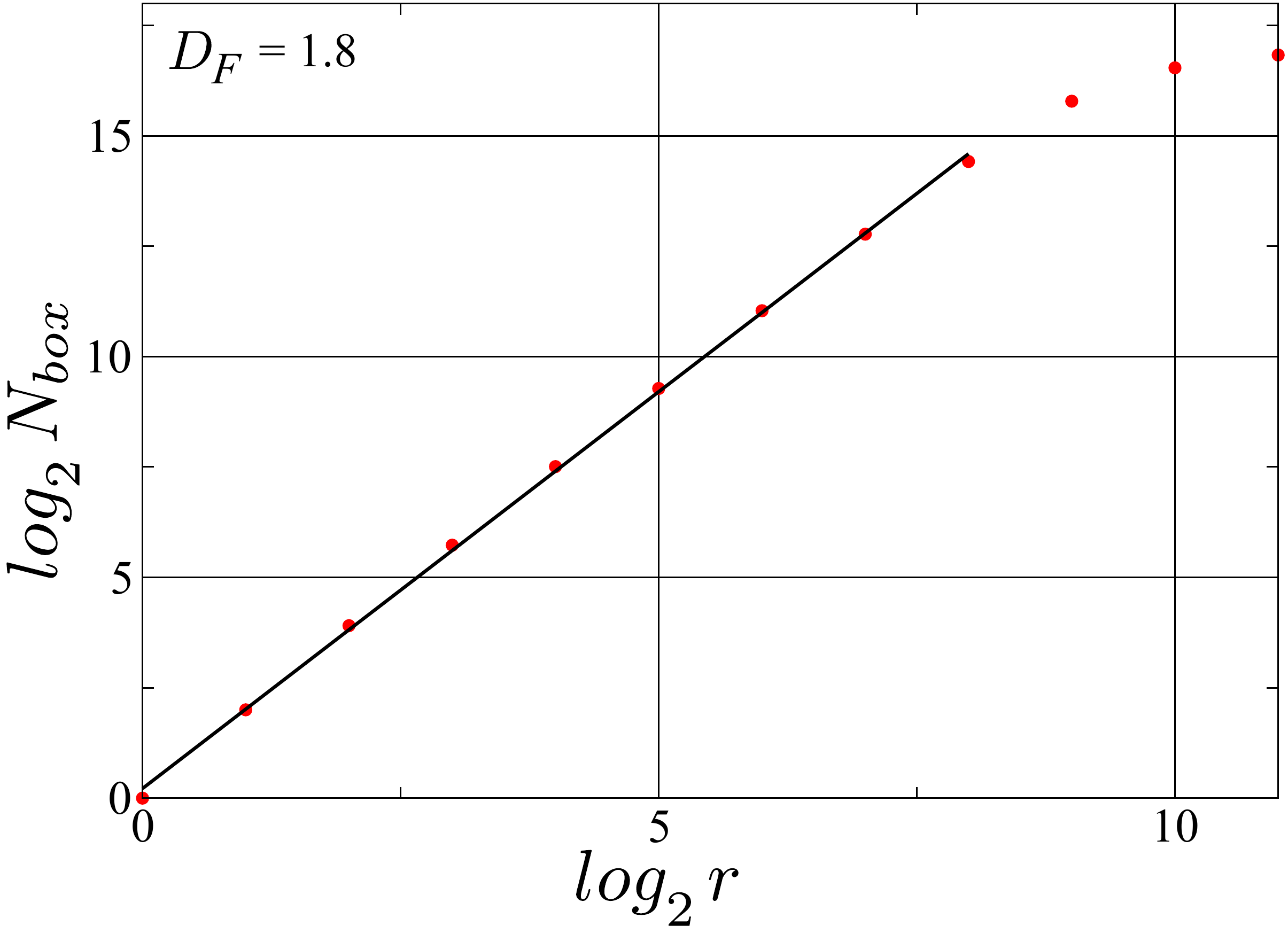}
\end{center}
\caption{Determination of the Minkowski (or box-counting) dimension for 
the attractor illustrated in Fig.~\ref{fig_chaos}. 
$N_{box}$ denotes the number of squares occupied with 
at least one point of the trajectory of consecutive pairs 
of spike intervals, when a grid of $2^r\times 2^r$ squares 
is laid upon the attracting set shown in 
Fig.~\ref{fig_chaos}. The fractal dimension
$D_F=log(N_{box})/log(r)$ is then $\sim$1.8. A
time series with $N_{spikes}=128000$ spikes for the same drifting 
neuron in the same D+S state has been used. 
$\log_2(N_{box})$ saturates at
$\log_2(N_{spikes})\approx 16.97$, observing that the linear 
range can be expanded further by increasing the number of 
spikes, albeit with an high cost in simulation time. Finally the
resolution of the method is limited by the spike width.
}
\label{fig_fractal_dim}
\end{figure}

We have determined the fractal dimension of
the attracting set of pairs of spike intervals
in the mixed phase by overlaying the attractor 
with a grid of $2^r\times2^ r$ squares. For this 
calculation we employed a longer simulation with 
$N_{spikes} = 128000$. The resulting 
box count, presented in Fig~\ref{fig_fractal_dim}, 
yields a Minkowski or box-counting dimension 
$D_F\approx1.8$, embeded in the 2D space of the plot, 
confirming such that the drifting neurons in 
the D+S phase spikes indeed chaotically. As a comparison, 
a limit cycle in this space, has a $D_F$ of $1$. 
While we present here the result for one particular neuron, 
the same holds for every drifting neuron in this state, 
albeit with slightly different fractal dimension values.
We note that the such determined fractal dimension
is not the one of the full attractor in $d=2N$ phase space, 
for which tuples of $2N$ consecutive inter-spike intervals
would need to be considered \cite{takens1981detecting,ding1993estimating}.
Our point here is that a non-integer result for the single 
neuron $D_F$ strongly indicates that the full attractor 
(in the $d$-dimensional phase space) is chaotic.\\

We believe that the chaotic state arising in the mixed
D+S state may be understood in analogy to
the occurrence of chaos in the periodically driven 
pendulum \cite{d1982chaotic}. A drifting neuron 
with a coupling constant $K$ in the D+S does indeed 
receive two types of inputs to its conductance, 
compare Eq.~(\ref{eq_g}), with the first input being 
constant (resulting from the firing of the other 
drifting neurons) and with the second input being 
periodic. The frequency $r_{syn}$ of the periodic 
driving will then be strictly smaller than the 
natural frequency $r_K$ of the drifting neuron
as resulting from the constant input (compare
Fig.~\ref{fig_examples}). It is known from the
theory of driven oscillators \cite{d1982chaotic}
that the oscillator may not be able to synchronize
with the external frequency, here $r_{syn}$,
when the frequency ratio $r_{syn}/r_K$ is 
small enough and the relative strength of 
the driving not too strong.\\

\begin{figure}[!t]
\begin{center}
\includegraphics[width=\textwidth]{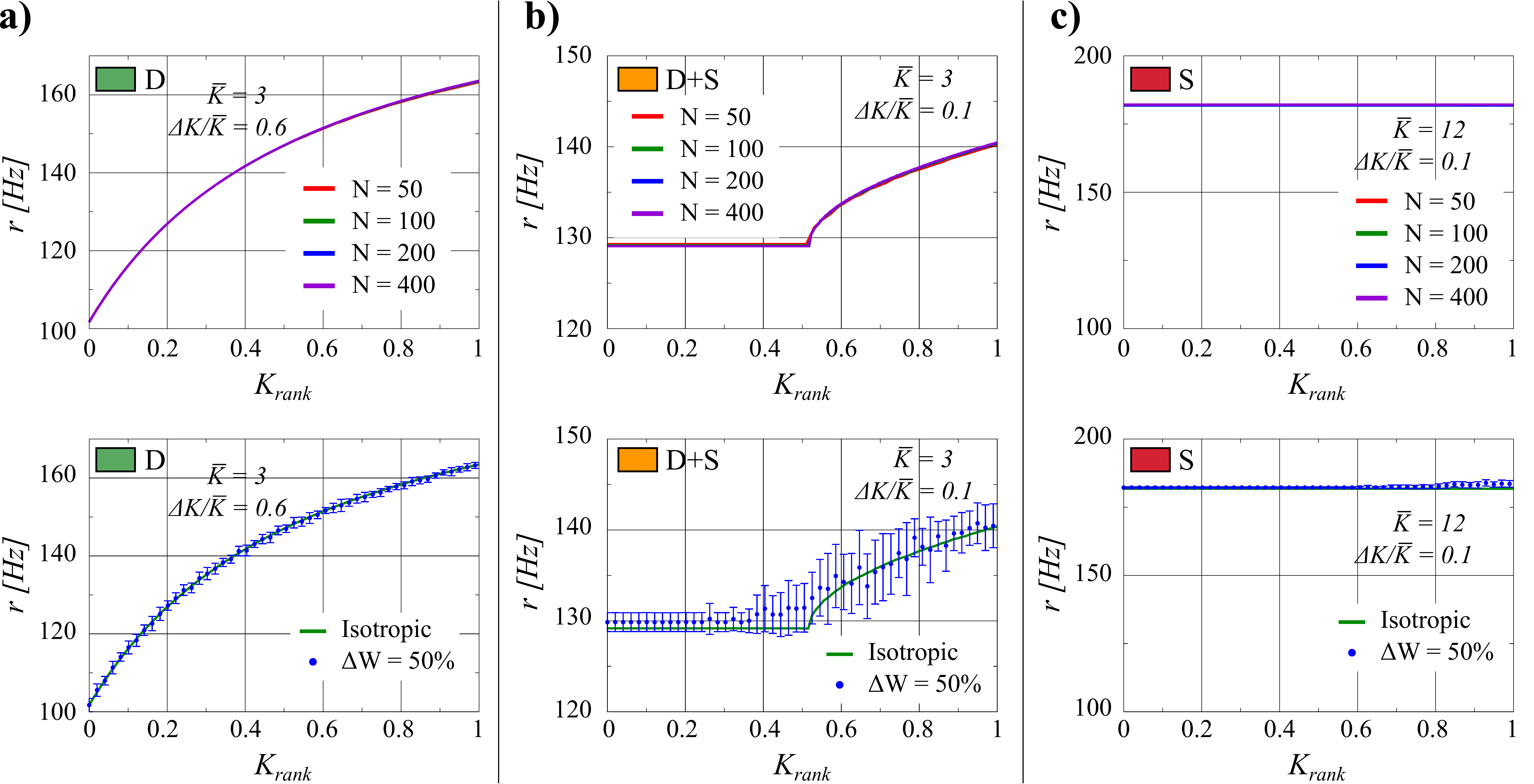}
\end{center}
\caption{As in Fig.~\ref{fig_examples}, the firing rate 
of each neuron in the network is presented as a function 
of the neuron's relative rank in $K$ (from smaller to larger). 
\textbf{a)} $\bar{K} = 3.0$, $\Delta K/\bar{K} = 0.6$ (fully drifting:
D).
\textbf{b)} $\bar{K} = 3.0$, $\Delta K/\bar{K} = 0.1$ (partially
drifting and synchronized: D+S).
\textbf{c)} $\bar{K} = 12.0$, $\Delta K/\bar{K} = 0.1$
(fully synchronized: S).
\textit{Top}: Comparison for several network sizes $N$. 
\textit{Bottom}: The numerical result for a network with
a uniform connectivity matrix (red line) in comparison
to a network in which the elements of the connectivity 
matrix are allowed to vary within $50\%$ up or down from 
unity (Blue points. The error bars represent the standard 
deviation of twenty realizations of the weight matrix).
}
\label{fig_robustness}
\end{figure}

\subsubsection{Robustness}

In order to evaluate the robustness and the 
generality of the results here presented, we 
have evaluated the effects occurring when changing 
the size of the network and when allowing for variability in the 
connectivity matrix $w_{ij}$. We have also considered an adiabatically 
increasing external input, as well as Gaussian noise.\\

In Fig.~\ref{fig_robustness} (top half), the effect on the rate 
distribution of the network size is evaluated. Sizes of 
$N=50,~100,~200$, and $400$ have been employed. 
We observe that the plots overlap
within the precision of the simulations. 
This result is on the one hand a consequence
of the scaling $K_i\sim 1/N$ for the overall strength of 
the afferent links and, on the other hand, of the regularity 
in firing discussed in Sect.~\ref{sec_time_structure}. 
The neural activity is driven by the mean field $\bar{r}(t)$
which is constant, in the thermodynamic limit $N\to\infty$, in 
the fully drifting state and non-constant but smooth (apart
from an initial jump) in the synchronized states. Fluctuations 
due to finite network sizes are already small for 
$N \approx 100$, as employed for our simulations, justifying 
this choice.\\

In the previous sections, we considered the uniform connectivity 
matrix described by (\ref{eq_w}). This allowed us to formulate 
the problem in terms of a mean-field coupling. We now analyze 
the robustness of the states found when a certain degree of 
variability is present in the weight matrix, viz when 
an extra variability term $\eta$ is present:
\begin{equation}
w_{ij} = 1 + \eta, \quad\quad
\eta \mathrm{\ random}, \quad\quad
i\ne j~.
\label{eq_w_eta}
\end{equation}
Here we consider $\eta$ to be drawn from a flat distribution 
with zero mean and a width $\Delta W$. Tests with 
$\Delta W = 10\%$, $20\%$, and $50\%$ were performed. 
In Fig.~\ref{fig_robustness} (lower half), the results for 
$\Delta W = 50\%$ are presented. We observe that the 
fully drifting state is the least affected by the 
variability in the weight matrix. On the other hand, 
the influence of variable weight matrices becomes
more evident when the state is partially
or fully synchronized, with the separation between 
the locked and the drifting neurons becoming less 
pronounced in the case of partial synchronization 
(lower panel of Fig.~\ref{fig_robustness}{\bf b)}). 
The larger standard deviation evident for larger values 
of $K_{rank}$ in the lower 
panel of Fig.~\ref{fig_robustness}{\bf c)} indicates the 
presence of drifting states in some of the ensemble
realizations of weight matrices.\\

\begin{figure}[!t]
\begin{center}
\includegraphics[width= 0.5 \textwidth]{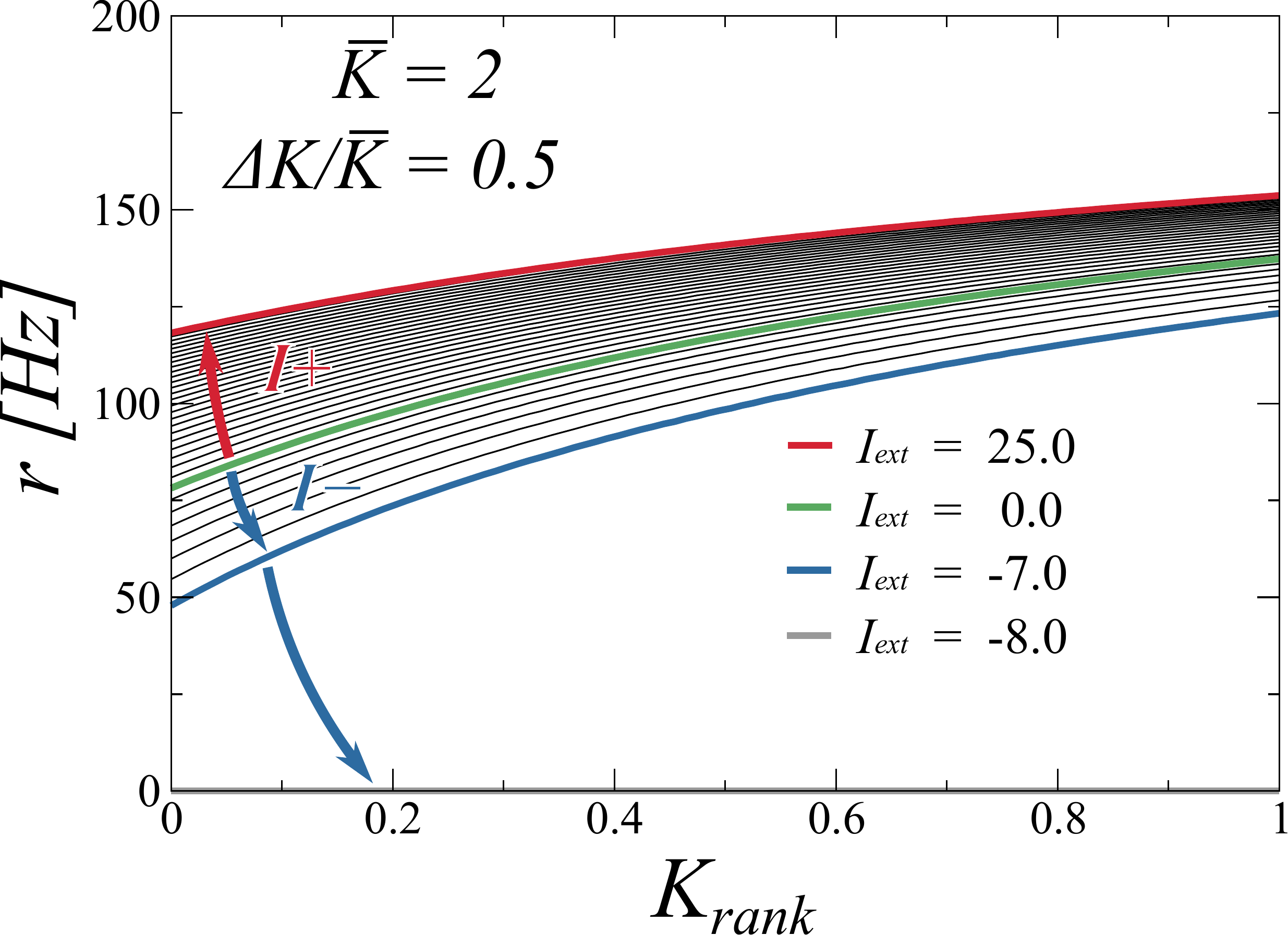}
\end{center}
\caption{Effect of an external input (either positive 
or negative), on the neural firing rates. The input is 
either increased (blue arrow) or decreased (red arrow)
from zero (green curve) adiabatically. The drifting state
remains stable for a wide range of inputs, with the
activity disappearing only for $I_{ext}<-7.0$ (grey curve,
coinciding with the $x$-axis).
}
\label{fig_stability}
\end{figure}

Finally, we test the robustness of the drifting state 
when perturbed with an external stimulus. To determine 
the stability of the state, we adiabatically increase 
the external stimulus $I_{ext}$ and compute the firing rate 
as a function of the rank for several values of $I_{ext}$. We 
do two excursions, one for positive values of $I_{ext}$ and 
another one for negative values. These plots are presented 
in Fig.~\ref{fig_stability}. We observe that the firing 
rates evolve in a continuous fashion, indicating that the drifting 
state is indeed stable. While positive inputs push the 
system to saturation, negative inputs reduce the average 
rate. We find, as is to be expected, that a large enough 
negative input makes the system silent. As a final test 
(not shown here), we have perturbed the system with 
random Gaussian uncorrelated noise, observing that 
the found attractors are all robust with respect to 
this type of noise as well.\\

\section{Discussion}

In the present work we have studied a network of excitable 
units, consisting exclusively of excitatory neurons. In absence of 
external stimulus, the network is only able to remain active 
through its own activity, in a self-organized fashion. Below 
a certain average coupling strength of the network the 
activity dies out, whereas, if the average coupling is 
strong enough, the excitable units will collectively behave 
as pulse-coupled oscillators.\\

We have shown how the variability of coupling 
strengths determines the synchronization characteristics 
of the network, ranging from fully asynchronous, to fully 
synchronous activity. Interestingly, this variability, 
together with the neurons' refractoriness, is enough to 
keep the neural activity from exploding.\\ 

While we have initially assumed a purely mean field coupling 
(by setting all the synaptic weights $w_{ij}=1$), only 
regulating the intensity with which a neuron integrates 
the mean field with the introduction of a scaling constant 
$K_i$, we have later shown how the here found states also 
survive when we allow the $w_{ij}s$ to individually vary 
up or down by up to a $50\%$ value. We have also shown how 
the variability in coupling strength makes the asynchronous 
or drifting state extremely robust with respect to strong 
homogeneous external inputs.\\

Finally, we have studied the time structure of spikes in 
the different dynamical states observed. It is in the time
domain that we find the main difference with natural neural 
networks. Spiking in real neurons is usually irregular, 
and it is often modeled as Poissonian, whereas in our 
network we found a very high degree of regularity, even in the 
asynchronous state. Only in the partially synchronous state 
we found a higher degree of variability, as a result from
chaotic behavior. We have determined the fractal dimension
of the respective strange attractor in the space of
pairs of consecutive interspike intervals, finding 
fractional values of roughly $1.8$ for the different 
neurons in the state.\\

While it has been often stated that inhibition is a 
necessary condition for bounded and uncorrelated activity, 
we have show here that uncorrelated aperiodic (and even chaotic) 
activity can be obtained with a network of excitatory-only 
connections, in a stable fashion and without external input. 
We are of course aware that the firing rates obtained 
in our simulations are high compared to \textit{in vivo} activity 
levels and that the degree of variability in the time domain of 
spikes is far from Poissonian. We have however incorporated in 
this work only variability of the inter-neural connectivity,
keeping other neural properties (such as the neural intrinsic 
parameters) constant and homogeneous. In this sense, it would be 
interesting to study in future work, how intrinsic and synaptic 
plasticity \cite{triesch2007synergies}, modify these statistics, 
incorporating plasticity in terms of interspike-times 
\cite{clopath2010connectivity, echeveste2015two}, and in terms 
of neural rates \cite{hyvarinen1998independent, bienenstock1982theory, 
echeveste2015objective}. Here, instead of trying to 
reproduce the detailed connection statistics in the brain, 
which would in any case never be realistic without inhibitory 
neurons, we have shown how a minimal variability model in 
terms of non uniform link matrices is able to give rise 
to asynchronous spiking states, even without inhibition. 
Our results indicate therefore that further studies are 
needed for an improved understanding of which features of 
the asynchronous spiking state depend essentially on
inhibition, and which do not.\\

We have shown here that autonomous activity
(sustained even in the absence of external inputs)
may arise in networks of coupled excitable units, viz
for units which are not intrinsically oscillating. We
have also proposed a new tool to study the appearance 
of chaos in spiking neural networks by applying a box 
counting method to consecutive pairs of inter-spike 
intervals from a single unit. This tool
is readily applicable both to experimental data
and to the results of theory simulations in general.

\section{Acknowledgments}

The support of the 
German Science Foundation (DFG) and the 
German Academic Exchange Service (DAAD) are
acknowledged.

\bibliographystyle{unsrt}
\bibliography{Spiking_Networks}

\begin{thebibliography}{10}

\bibitem{pikovsky2015dynamics}
Arkady Pikovsky and Michael Rosenblum.
\newblock Dynamics of globally coupled oscillators: Progress and perspectives.
\newblock {\em Chaos: An Interdisciplinary Journal of Nonlinear Science},
  25(9):097616, 2015.

\bibitem{winfree1967biological}
Arthur~T Winfree.
\newblock Biological rhythms and the behavior of populations of coupled
  oscillators.
\newblock {\em Journal of theoretical biology}, 16(1):15--42, 1967.

\bibitem{peskin1975mathematical}
Charles~S Peskin.
\newblock {\em Mathematical aspects of heart physiology}.
\newblock Courant Institute of Mathematical Sciences, New York University,
  1975.

\bibitem{kuramoto1975self}
Yoshiki Kuramoto.
\newblock Self-entrainment of a population of coupled non-linear oscillators.
\newblock In {\em International symposium on mathematical problems in
  theoretical physics}, pages 420--422. Springer, 1975.

\bibitem{buck1988synchronous}
John Buck.
\newblock Synchronous rhythmic flashing of fireflies. ii.
\newblock {\em Quarterly review of biology}, pages 265--289, 1988.

\bibitem{gros2010complex}
C.~Gros.
\newblock {\em Complex and adaptive dynamical systems: A primer}.
\newblock Springer Verlag, 2010.

\bibitem{levnajic2010phase}
Zoran Levnaji{\'c} and Arkady Pikovsky.
\newblock Phase resetting of collective rhythm in ensembles of oscillators.
\newblock {\em Physical Review E}, 82(5):056202, 2010.

\bibitem{ashwin2007dynamics}
Peter Ashwin, G{\'a}bor Orosz, John Wordsworth, and Stuart Townley.
\newblock Dynamics on networks of cluster states for globally coupled phase
  oscillators.
\newblock {\em SIAM Journal on Applied Dynamical Systems}, 6(4):728--758, 2007.

\bibitem{dorfler2014synchronization}
Florian D{\"o}rfler and Francesco Bullo.
\newblock Synchronization in complex networks of phase oscillators: A survey.
\newblock {\em Automatica}, 50(6):1539--1564, 2014.

\bibitem{golomb1992clustering}
D~Golomb, David Hansel, B~Shraiman, and Haim Sompolinsky.
\newblock Clustering in globally coupled phase oscillators.
\newblock {\em Physical Review A}, 45(6):3516, 1992.

\bibitem{abbott1993asynchronous}
LF~Abbott and Carl van Vreeswijk.
\newblock Asynchronous states in networks of pulse-coupled oscillators.
\newblock {\em Physical Review E}, 48(2):1483, 1993.

\bibitem{strogatz1993coupled}
Steven~H Strogatz, Ian Stewart, et~al.
\newblock Coupled oscillators and biological synchronization.
\newblock {\em Scientific American}, 269(6):102--109, 1993.

\bibitem{mirollo1990synchronization}
Renato~E Mirollo and Steven~H Strogatz.
\newblock Synchronization of pulse-coupled biological oscillators.
\newblock {\em SIAM Journal on Applied Mathematics}, 50(6):1645--1662, 1990.

\bibitem{hanson1978comparative}
Frank~E Hanson.
\newblock Comparative studies of firefly pacemakers.
\newblock In {\em Federation proceedings}, volume~37, pages 2158--2164, 1978.

\bibitem{buzsaki2004neuronal}
Gy{\"o}rgy Buzs{\'a}ki and Andreas Draguhn.
\newblock Neuronal oscillations in cortical networks.
\newblock {\em Science}, 304(5679):1926--1929, 2004.

\bibitem{velazquez2007phase}
JL~Perez Velazquez, RF~Galan, L~Garcia Dominguez, Y~Leshchenko, S~Lo, J~Belkas,
  and R~Guevara Erra.
\newblock Phase response curves in the characterization of epileptiform
  activity.
\newblock {\em Physical Review E}, 76(6):061912, 2007.

\bibitem{kuramoto1991collective}
Yoshiki Kuramoto.
\newblock Collective synchronization of pulse-coupled oscillators and excitable
  units.
\newblock {\em Physica D: Nonlinear Phenomena}, 50(1):15--30, 1991.

\bibitem{izhikevich1999weakly}
Eugene~M Izhikevich.
\newblock Weakly pulse-coupled oscillators, fm interactions, synchronization,
  and oscillatory associative memory.
\newblock {\em Neural Networks, IEEE Transactions on}, 10(3):508--526, 1999.

\bibitem{strogatz2000kuramoto}
Steven~H Strogatz.
\newblock From {Kuramoto} to {Crawford}: exploring the onset of synchronization
  in populations of coupled oscillators.
\newblock {\em Physica D: Nonlinear Phenomena}, 143(1):1--20, 2000.

\bibitem{sonnenschein2014cooperative}
Bernard Sonnenschein, Thomas K~DM Peron, Francisco~A Rodrigues, J{\"u}rgen
  Kurths, and Lutz Schimansky-Geier.
\newblock Cooperative behavior between oscillatory and excitable units: the
  peculiar role of positive coupling-frequency correlations.
\newblock {\em The European Physical Journal B}, 87(8):1--11, 2014.

\bibitem{burkitt2006review}
Anthony~N Burkitt.
\newblock A review of the integrate-and-fire neuron model: I. homogeneous
  synaptic input.
\newblock {\em Biological cybernetics}, 95(1):1--19, 2006.

\bibitem{shadlen1994noise}
Michael~N Shadlen and William~T Newsome.
\newblock Noise, neural codes and cortical organization.
\newblock {\em Current opinion in neurobiology}, 4(4):569--579, 1994.

\bibitem{amit1997model}
Daniel~J Amit and Nicolas Brunel.
\newblock Model of global spontaneous activity and local structured activity
  during delay periods in the cerebral cortex.
\newblock {\em Cerebral cortex}, 7(3):237--252, 1997.

\bibitem{sanchez2000cellular}
Maria~V Sanchez-Vives and David~A McCormick.
\newblock Cellular and network mechanisms of rhythmic recurrent activity in
  neocortex.
\newblock {\em Nature neuroscience}, 3(10):1027--1034, 2000.

\bibitem{haider2006neocortical}
Bilal Haider, Alvaro Duque, Andrea~R Hasenstaub, and David~A McCormick.
\newblock Neocortical network activity in vivo is generated through a dynamic
  balance of excitation and inhibition.
\newblock {\em The Journal of neuroscience}, 26(17):4535--4545, 2006.

\bibitem{van1996chaos}
Carl van Vreeswijk and Haim Sompolinsky.
\newblock Chaos in neuronal networks with balanced excitatory and inhibitory
  activity.
\newblock {\em Science}, 274(5293):1724--1726, 1996.

\bibitem{kumar2008high}
Arvind Kumar, Sven Schrader, Ad~Aertsen, and Stefan Rotter.
\newblock The high-conductance state of cortical networks.
\newblock {\em Neural computation}, 20(1):1--43, 2008.

\bibitem{vogels2005signal}
Tim~P Vogels and Larry~F Abbott.
\newblock Signal propagation and logic gating in networks of integrate-and-fire
  neurons.
\newblock {\em The Journal of neuroscience}, 25(46):10786--10795, 2005.

\bibitem{stefanescu2008low}
Roxana~A Stefanescu, Viktor~K Jirsa, et~al.
\newblock A low dimensional description of globally coupled heterogeneous
  neural networks of excitatory and inhibitory neurons.
\newblock {\em PLoS Comput. Biol}, 4(11):e1000219, 2008.

\bibitem{hansel2001existence}
D~Hansel and G~Mato.
\newblock Existence and stability of persistent states in large neuronal
  networks.
\newblock {\em Physical Review Letters}, 86(18):4175, 2001.

\bibitem{brunel2000dynamics}
Nicolas Brunel.
\newblock Dynamics of sparsely connected networks of excitatory and inhibitory
  spiking neurons.
\newblock {\em Journal of computational neuroscience}, 8(3):183--208, 2000.

\bibitem{kuramoto2002coexistence}
Yoshiki Kuramoto and Dorjsuren Battogtokh.
\newblock Coexistence of coherence and incoherence in nonlocally coupled phase
  oscillators.
\newblock {\em arXiv preprint cond-mat/0210694}, 2002.

\bibitem{alonso2011average}
Leandro~M Alonso and Gabriel~B Mindlin.
\newblock Average dynamics of a driven set of globally coupled excitable units.
\newblock {\em Chaos: An Interdisciplinary Journal of Nonlinear Science},
  21(2):023102, 2011.

\bibitem{abrams2004chimera}
Daniel~M Abrams and Steven~H Strogatz.
\newblock Chimera states for coupled oscillators.
\newblock {\em Physical review letters}, 93(17):174102, 2004.

\bibitem{miritello2009central}
Giovanna Miritello, Alessandro Pluchino, and Andrea Rapisarda.
\newblock Central limit behavior in the {Kuramoto} model at the
  \textquotedblleft edge of chaos\textquotedblright.
\newblock {\em Physica A: Statistical Mechanics and its Applications},
  388(23):4818--4826, 2009.

\bibitem{angulo2014stable}
David Angulo-Garcia and Alessandro Torcini.
\newblock Stable chaos in fluctuation driven neural circuits.
\newblock {\em Chaos, Solitons \& Fractals}, 69:233--245, 2014.

\bibitem{lindner2004interspike}
Benjamin Lindner.
\newblock Interspike interval statistics of neurons driven by colored noise.
\newblock {\em Physical Review E}, 69(2):022901, 2004.

\bibitem{perkel1967neuronal_I}
Donald~H Perkel, George~L Gerstein, and George~P Moore.
\newblock Neuronal spike trains and stochastic point processes: I. the single
  spike train.
\newblock {\em Biophysical journal}, 7(4):391, 1967.

\bibitem{perkel1967neuronal_II}
Donald~H Perkel, George~L Gerstein, and George~P Moore.
\newblock Neuronal spike trains and stochastic point processes: Ii.
  simultaneous spike trains.
\newblock {\em Biophysical journal}, 7(4):419, 1967.

\bibitem{farkhooi2009serial}
Farzad Farkhooi, Martin~F Strube-Bloss, and Martin~P Nawrot.
\newblock Serial correlation in neural spike trains: Experimental evidence,
  stochastic modeling, and single neuron variability.
\newblock {\em Physical Review E}, 79(2):021905, 2009.

\bibitem{chacron2004noise}
Maurice~J Chacron, Benjamin Lindner, and Andr{\'e} Longtin.
\newblock Noise shaping by interval correlations increases information
  transfer.
\newblock {\em Physical review letters}, 92(8):080601, 2004.

\bibitem{d1982chaotic}
D~d'Humieres, MR~Beasley, BA~Huberman, and A~Libchaber.
\newblock Chaotic states and routes to chaos in the forced pendulum.
\newblock {\em Physical Review A}, 26(6):3483, 1982.

\bibitem{takens1981detecting}
Floris Takens.
\newblock Detecting strange attractors in turbulence.
\newblock In {\em Dynamical systems and turbulence, Warwick 1980}, pages
  366--381. Springer, 1981.

\bibitem{ding1993estimating}
Mingzhou Ding, Celso Grebogi, Edward Ott, Tim Sauer, and James~A Yorke.
\newblock Estimating correlation dimension from a chaotic time series: when
  does plateau onset occur?
\newblock {\em Physica D: Nonlinear Phenomena}, 69(3):404--424, 1993.

\bibitem{triesch2007synergies}
Jochen Triesch.
\newblock Synergies between intrinsic and synaptic plasticity mechanisms.
\newblock {\em Neural Computation}, 19(4):885--909, 2007.

\bibitem{clopath2010connectivity}
Claudia Clopath, Lars B{\"u}sing, Eleni Vasilaki, and Wulfram Gerstner.
\newblock Connectivity reflects coding: a model of voltage-based stdp with
  homeostasis.
\newblock {\em Nature neuroscience}, 13(3):344--352, 2010.

\bibitem{echeveste2015two}
Rodrigo Echeveste and Claudius Gros.
\newblock Two-trace model for spike-timing-dependent synaptic plasticity.
\newblock {\em Neural computation}, 2015.

\bibitem{hyvarinen1998independent}
Aapo Hyv{\"a}rinen and Erkki Oja.
\newblock Independent component analysis by general nonlinear hebbian-like
  learning rules.
\newblock {\em Signal Processing}, 64(3):301--313, 1998.

\bibitem{bienenstock1982theory}
Elie~L Bienenstock, Leon~N Cooper, and Paul~W Munro.
\newblock Theory for the development of neuron selectivity: orientation
  specificity and binocular interaction in visual cortex.
\newblock {\em The Journal of Neuroscience}, 2(1):32--48, 1982.

\bibitem{echeveste2015objective}
Rodrigo Echeveste and Claudius Gros.
\newblock An objective function for self-limiting neural plasticity rules.
\newblock In {\em European Symposium on Artificial Neural Networks,
  Computational Intelligence and Machine Learning. Bruges (Belgium), 22-24
  April 2015}, number ESANN 2015 proceedings. i6doc.com publ., 2015.

\end{thebibliography}

\end{document}